\documentclass{article}

\usepackage{authblk}
\usepackage{mathptmx}

\usepackage{comment}
\usepackage{amsmath,amssymb,amssymb} 
\usepackage{cite,array}
\usepackage{amsmath,amssymb,amsfonts}
\usepackage{algorithmic}
\usepackage{graphicx}
\usepackage{textcomp,multirow}
\usepackage{cases}
\usepackage[normalem]{ulem,mathtools}
\usepackage{flushend}

\title{A Model-based Approach for Glucose Control via Physical Activity \footnote{This manuscript is the arxiv version of the work in \cite{DePaolaMIE2025}}}

\author{
    Pierluigi Francesco De Paola$^{1,2,3}$, Alessandro Borri$^{3}$,
    Alessia Paglialonga$^{**,2}$, Pasquale Palumbo $^{4}$, Fabrizio Dabbene$^{2}$
}

\affil{
\small 
    $^1$Politecnico di Bari, Bari, Italy\\
    $^2$Consiglio Nazionale delle Ricerche, Istituto di Elettronica e di Ingegneria dell’Informazione e delle
Telecomunicazioni (CNR-IEIIT), Turin, Italy\\
    $^3$Consiglio Nazionale delle Ricerche, Istituto di  Analisi dei Sistemi ed Informatica (CNR-IASI), Rome, Italy\\
    $^4${Dept. of Biotechnologies and Biosciences, University of Milano-Bicocca, Milan, Italy}\\
    $^{**}$Corresponding author: alessia.paglialonga@cnr.it
}

\date{}
\begin{document}

\maketitle

%\pretitle{Pretitle}
%\title{Closed-loop Control of Long-term Diabetes Progression via Physical Activity: a Model-based Approach}

\begin{abstract}
The role played by physical activity in slowing down the progression of type-2 diabetes is well recognized. However, except for general clinical guidelines, quantitative real-time estimates of the recommended amount of physical activity, based on the evolving individual conditions, are {still missing} in the literature. 
The aim of this work is to provide a control-theoretical formulation of the exercise encoding all the exercise-related features (intensity, duration, period). Specifically, we design a feedback law in terms of recommended physical activity, following a model predictive control 
approach, based on a widespread compact diabetes progression model, suitably modified to account for the long-term effects of regular exercise.  
Preliminary simulations show 
promising results, well aligned with clinical evidence. These findings can be the basis for further validation of the control law on high-dimensional diabetes progression models to ultimately translate the predictions of the controller into meaningful recommendations. 
\end{abstract}

\section{Introduction}
{Type} 2 diabetes (T2D) is a chronic disease that is becoming more and more challenging worldwide, bearing a range of significant complications and % its rising {incidence}. Estimations forecast 
a growing burden {on the} healthcare systems, with substantial social and economic implications \cite{ZimmetEtAl2001}.
Evidence suggests that T2D can be prevented or significantly slowed down through lifestyle interventions, for example via regular physical activity.
Nevertheless, there is a lack {of} mathematical models suitably describing the physiological machinery mediating the effect of physical activity on T2D course.
Within the framework of the research line named "Artificial Pancreas", control approaches have been focusing on  short-term, model-based, glucose control \cite{BorriEtAl2021,IncremonaEtAl2018,BorriEtAl2017} via insulin administration and no model-based control techniques so far have leveraged physical activity management for glucose control and T2D prevention in the long term. 
The aim of the present study is to provide a control-theoretical formulation of the exercise by introducing a novel control-driven representation to design a model predictive control (MPC) \cite{Allgower2012} on a compact model of T2D progression. Specifically, the model by Topp et al. \cite{ToppEtAl2000} was suitably modified to integrate the effect of physical activity mediated by Interleukin-6, as formulated in our previous works \cite{DePaolaEtAl,de2024novel}.
Hence, to the best of our knowledge, in this contribution we provide an original model-based approach for long-term glucose control via physical activity management. 

\section{Methods}
\label{sec:SimulationResults}
The model here used is {a  modification to the one originally proposed by Topp \textit{et al.} ~\cite{ToppEtAl2000}}. It describes the evolution of glucose/insulin and the state variables that control their homeostasis on a long period (months) and the integral effect of physical activity (modeled by the newly introduced state variable $V_l$). The main equations and variables are described in Table \ref{model}, whereas a detailed description is reported in \cite{ToppEtAl2000,de2024novel}.
\begin{table}[h!]
    \centering
    \begin{tabular}{|c|p{6cm}|}
        \hline
        \textbf{Equation} & \textbf{Description of the variables} \\
        \hline
        $\dot{G} = R_0 + \left(E_{g0} + S_I I\right) G$ & $G$, plasma glucose concentration [mg/dl] \\
        \hline
        $\dot{I} = \beta \sigma \frac{G^2}{\alpha + G^2} - k I$ & $I$, serum insulin concentration [$\mu$U/ml] \\
        \hline
        $\dot{\beta} = (\bar{P} - \bar{A}) \beta$ & $\beta$, beta cell mass [mg]\\
        \hline
        $\dot{S_I} = -c(S_I - S_{I,\text{{target}}}) \left(1 - \zeta_{si} \frac{V_l}{k_{n,si} + V_l}\right)$ & $S_I$, insulin sensitivity [ml/$\mu$U/d] \\
        \hline
        $\dot{V_l} = \frac{SR}{K_{IL6}} \cdot u - k_s V_l$ & $V_l$, integral effect of physical activity (see \cite{DePaolaEtAl,de2024novel} for details) %IL-6 
        [(pg/dl)*min], representing the long-term effect of exercise\\
        \hline
        $\bar{P} = P(G) \cdot \psi_1(V_l),\quad \psi_1(V_l)=\left(1 + \frac{\zeta_{p} V_l^2}{k_{p}^2 + V_l^2}\right)$ & $\bar{P}$, beta cell mass proliferation [1/d] when including the effect of the exercise  \\
        \hline
        $\bar{A} = A(G) \cdot \psi_2(V_l),\quad \psi_2(V_l)=\left(1 - \frac{\zeta_{a} V_l^2}{k_{a}^2 + V_l^2}\right)$ & $\bar{A}$, beta cell mass apoptosis [1/d] when including the effect of the exercise\\
        \hline
        $P(G) = r_{1r} G - r_{2r} G^2$ & $P$, beta cell mass proliferation [1/d]  \\
        \hline
        $A(G) = d_0 - r_{1a} G + r_{2a} G^2$ & $A$, beta cell mass apoptosis [1/d]\\
        \hline
    \end{tabular}
    \caption{Model exploited in this work for the design of the model-based control algorithm }
    \label{model}
\end{table}

Glucose/insulin, beta-cell mass, and insulin sensitivity dynamics are inherited from \cite{ToppEtAl2000} and the beta-cell dynamics is modified with respect to the original work of \cite{ToppEtAl2000} to incorporate physical activity benefits. Specifically, while $\psi_1$ and $\psi_2$ were formally set to 1 in the original model, in this study we reformulate these variables using Hill functions of the $V_l$ state variable, similarly to our previous works \cite{DePaolaEtAl,de2024novel} to highlight the separate contribution of physical activity on beta-cell proliferation ${P}(G)$ and apoptosis ${A}(G)$.
Moreover, with respect to~\cite{ToppEtAl2000}, a general positive $S_{I\!,\text{\it target}}$ is introduced in  $S_I$ dynamics, as in \cite{DePaolaEtAl,de2024novel}, with the aim of avoiding the unrealistic, unbounded, beta-cell growth of the original model by Topp \textit{et al.} \cite{ToppEtAl2000}.
Finally, we model the additional effect of exercise in terms of improved insulin sensitivity $S_I$, widely known in the literature \cite{bird2017update},
by adding a factor related to physical activity as described in detail in our previous work \cite{de2024novel}. 
{For what concerns} the control input $u$ incorporating physical activity, it is defined starting from the original formulation in \cite{DePaolaEtAl,de2024novel} by removing the
fast dynamics to simplify the model-based control scheme, given the aim to control T2D progression in the long term. 
Specifically, in \cite{DePaolaEtAl,de2024novel} the 
variable $u$ is defined as a piecewise-constant input representing the exercise intensity, performed regularly with a given period $T$ in session of duration $\delta$.
In this work, to derive a control-theoretical formulation of the effect of the exercise on T2D progression, we introduce an equivalent constant input $u_{eq}$ in a compact form - representing the average effect of the physical activity 
- by distributing the effect of the exercise all over the period $T$: $u(t)=u_{eq} = \frac{\bar{u}\cdot\delta + 0\cdot (T-\delta)}{T} = \frac{\bar{u}\cdot \delta}{T}, \ t\in[0,T].$ 
Once  the control is computed, by keeping fixed the exercise intensity $\bar{u}$ and the period $T$ of the exercise sessions, by means of the inverse map $\delta =  \frac{ u_{eq} \cdot T}{\bar{u}}, \label{eq_inverse_map}$ it is possible to {translate} the information of a general exercise program encoded in $u_{eq}$ into precise recommendations in terms of duration of the exercise to be performed. The exercise program is updated accordingly in agreement with the suggestions provided by the controller, designed by means of an MPC formulation \cite{Allgower2012}. 
More formally, in our case the MPC problem finds the optimal control sequence $\{u^*_{eq,k}\}_{k\in\mathbb{N}}$, and is designed as follows:
\vspace{-15pt}
\setlength{\jot}{0pt} 
\begin{align}\label{eq:u_star}
    u^*_{eq,k} = &\arg\min_{u_{eq}\in[0,u_{eq}^{\max}]} \displaystyle\int_{kT}^{(k+N)T} (G(s)^2  +\lambda\cdot u_{eq}^2)\, ds \\
    \hspace{-3mm}\text{s.t.} & \quad \begin{aligned}[t]
        & \dot{x} = f(x(t),u(t)), \quad  t\geq 0, \\
        & u(t)= u^*_{eq,k}, \quad \qquad  t \in [kT,(k+1)T), \quad  k \in \mathbb{N}. \nonumber
        %\\
        %& 0 \leq u_{eq,k} \leq 3,
    \end{aligned}
\end{align}
The constraints account for the dynamics of the system  
and for the  bounds on $u_{eq}$. Specifically, we consider $u_{eq}^{\max}=3$    since, when fixing $u=60\%$ (moderate-to-vigorous intensity), $T=2$ days, the inverse map
translates the maximal allowed regime $u_{eq}=u_{eq}^{\max}=3$    into an equivalent overall duration of exercise very close to 400 minutes/week, that is the weekly duration at which the benefits of moderate-to-vigorous intensity exercise on diabetes prevention saturate \cite{boonpor2023dose}.
Assuming that the conditions predisposing to T2D arise at $t=0$,
we simulate the model with initial conditions $x(0)=[G(0) \quad  I(0) \quad \beta(0) \quad S_I(0) \quad  V_l(0)]^T=[100 \quad  10 \quad 300 \quad 0.72 \quad 0]^T$ both in the open loop case and in the controlled case, feeding the system with the prediction progressively updated by the controller.
The prediction window $N$ is set equal to $20$ and parameter $\lambda$ is set equal to $60$. Since T2D progression according to the model by Topp \textit{et al.} \cite{ToppEtAl2000} occurs tipically over a time span of one year, the same time frame is considered for our simulations.

\begin{figure*}[!h]
\centering
\begin{minipage}{0.49\textwidth}
    \centering
    \hspace*{-0.7 cm}
    \includegraphics[scale=0.365]{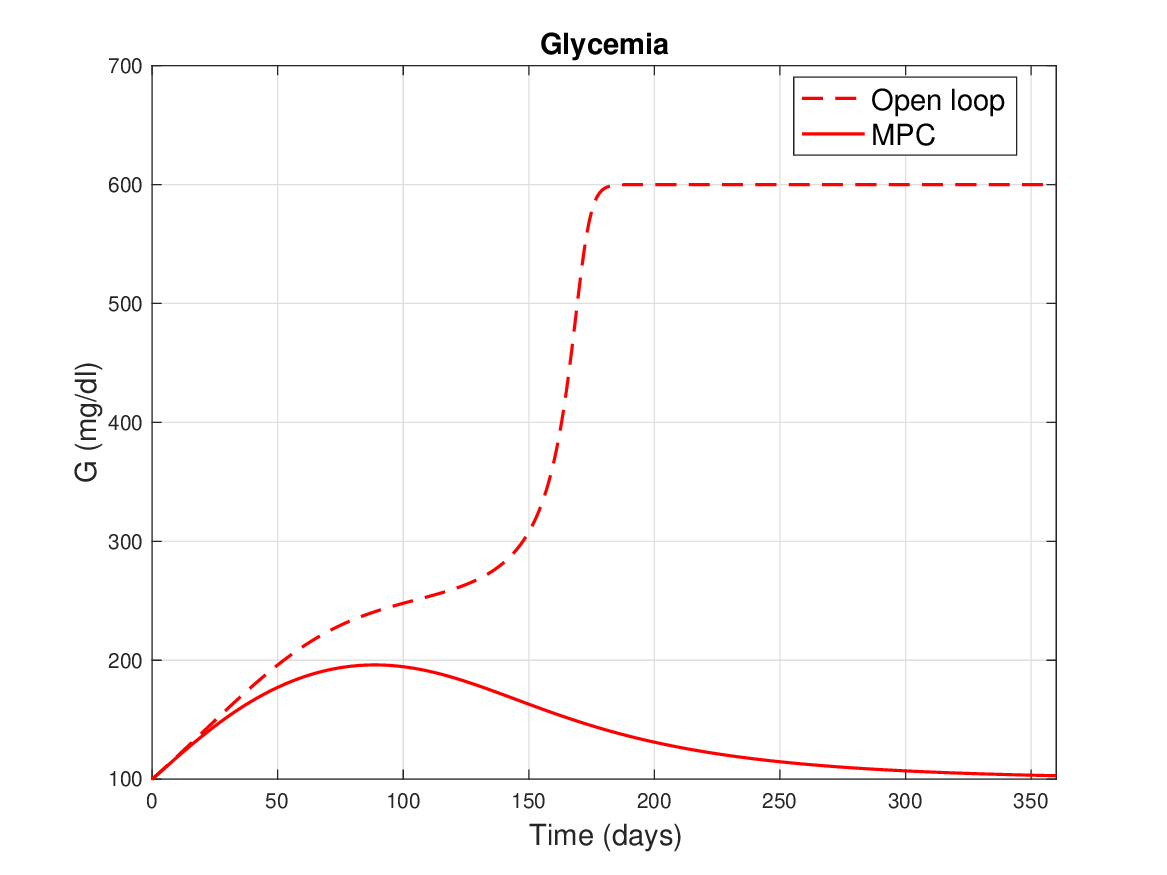}
    \caption{Basal glucose concentration as a function of time in the open loop case (dashed line) and in the controlled case (solid line).}
    \label{fig:G_basal}
\end{minipage}%
\hfill
\begin{minipage}{0.49\textwidth}
    \centering
    \hspace*{-0.45 cm} % Matching the same hspace for all figures
    \includegraphics[scale=0.365]{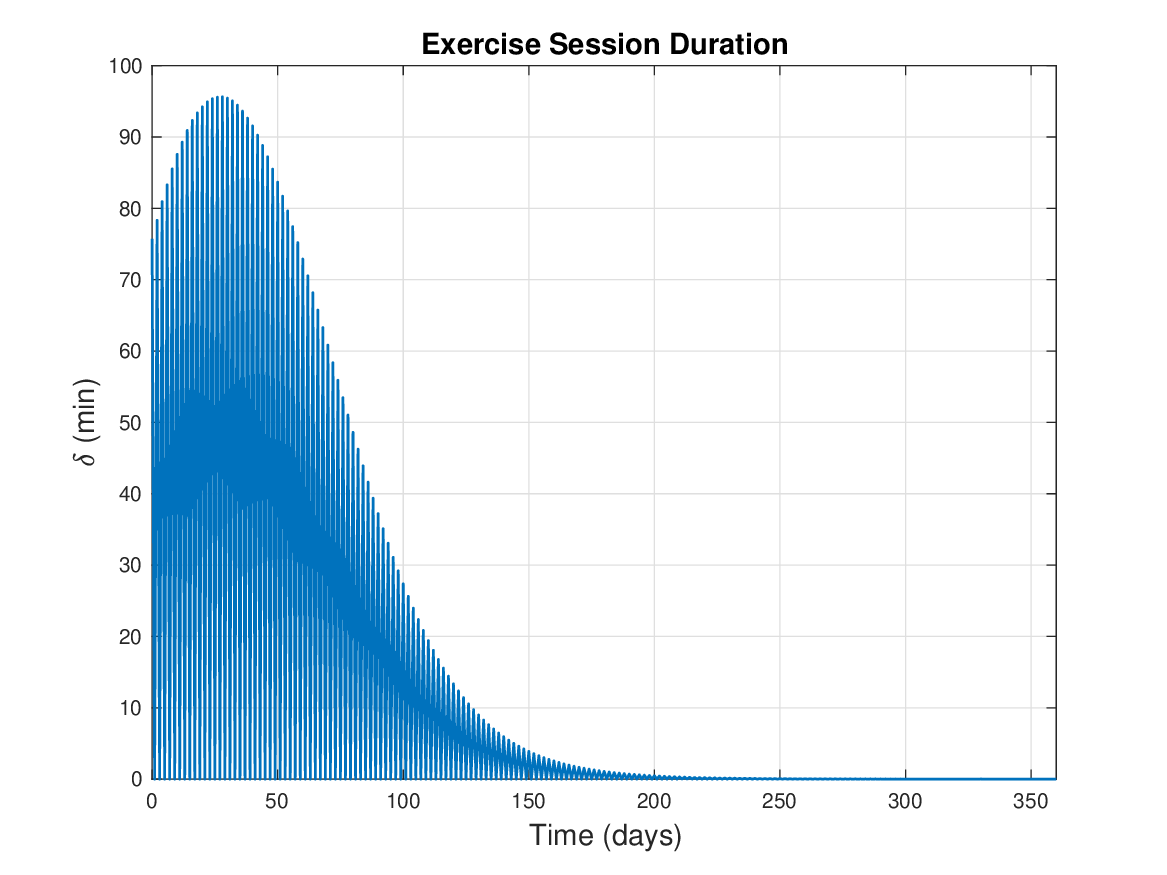}
    \caption{Recommended duration of single exercise session as a function of time computed by means of the inverse map in the MPC-controlled case ($ \bar{u} = 50\%, T= 2 $ days).}
    \label{fig:duration}
\end{minipage}
\end{figure*}

\section{Results}

Fig. \ref{fig:G_basal} shows the basal glucose concentration in the open-loop case (dashed line, no exercise) and in the MPC-controlled case (solid line, exercise determined by eq. \eqref{eq:u_star}). As it can be observed, in the open-loop case the system undergoes a severe T2D progression and the progressively rising glucose levels makes the course of the disease irreversible approximately from the 100th day, as beta cells fail the dynamical compensation because of the effects of glucotoxicity \cite{ToppEtAl2000}. As a result, the system reaches the hyperglycemic steady state (that is conventionally set at $G=600$ mg/dl in the Topp model\cite{ToppEtAl2000,case2024}). Conversely, in closed loop, with the MPC recommending the equivalent control input $u_{eq}$ as the amount of exercise to be performed, the system is able to delay the course of T2D and, in the long term, to reverse the progression and restore normal values for the basal glucose concentration, reaching the normo-glycemic steady-state (that is conventionally set at $G=100$ in the Topp's model\cite{ToppEtAl2000,case2024}). 
Fig. \ref{fig:duration} shows how the computed control law in terms of equivalent input ~$u_{eq}$, encoding the overall information about the exercise program, can be translated into a precise, time-varying recommendation on the amount of exercise to be performed. Specifically, by fixing the exercise intensity  (i.e., $\bar{u}=50\%$ simulating moderate-intensity exercise) and the period (i.e., $T=2$ days), by means of the inverse map, 
the optimal MPC input \eqref{eq:u_star} is converted into a precise recommendation on the duration of single exercise sessions. 
As it can be seen from the plot, in the early phases of T2D progression, the overall amount of exercise duration is close to $300$ min/week and it gradually decreases, the duration of the exercise being updated by the MPC controller accordingly with the progressively lower values of glucose.
\section{Discussion}
This work represents a proof of concept showing how a control law should be designed to leverage physical activity
to control T2D progression. Notably, our results are consistent with evidence in the literature concerning T2D prevention programs through lifestyle interventions. Indeed, to preserve beta-cell mass from degradation due to the progression of disease, a higher dose of exercise should be performed in the early stages of the diabetes course (Fig. \ref{fig:duration})\cite{ToppEtAl2000,DePaolaEtAl}. This also aligns with the fact that the benefits of physical activity may persist in the long term even after a discontinuation of the intervention \cite{uusitupa2003long}. 
Moreover, WHO general guidelines on T2D prevention suggest a minimum of $150$ min/week of moderate exercise intensity, with higher benefits expected with higher doses of exercise \cite{bull2020world,boonpor2023dose}, in agreement with the overall duration of the prediction of our controller, suggesting about 300 minutes/week of exercise in the early stage of T2D progression.
The preliminary results here shown are promising and future research is needed to further develop model-driven approaches for T2D prevention via physical activity. 
{However, this work shows also some limitations that should be overcome in future studies. Indeed, at this stage the model underlying the control law does not account for parameter variations that would allow to simulate patients inter-variability and does not account explicitly for meal intakes. Moreover, the model is quite simple in its state-space representation, involving only five state variables for the design of the control law. Leveraging a more complex model, including the  action of additional state variables, would allow to further account for a more detailed description of the physiological mechanisms describing the effect of physical activity on diabetes progression.Future developments are aimed at ovecoming these limitations, exploiting the extended version of our model \cite{DePaolaEtAl,de2024novel} to provide a quantitative assessment to the general, experience-driven, medical advice on exercise programs for T2D prevention~\cite{bull2020world} }
\vspace{-10pt}
\section{Acknowledgment}
{This work was supported in part by the European Union through the Project PRAESIIDIUM “Physics Informed Machine Learning-Based Prediction and Reversion of Impaired Fasting Glucose Management" (call HORIZON-HLTH-2022-STAYHLTH-02), Grant 101095672. Views and opinions expressed are however those of the authors only and do not necessarily reflect those of the European Union or the European Health and Digital Executive Agency (HADEA). Neither the European Union nor the HADEA can be held responsible for them. This work was carried out within the Italian National Ph.D. Program in Autonomous Systems (DAuSy), coordinated by Polytechnic of Bari, Italy. 
}
\bibliographystyle{vancouver}
\bibliography{Topp_Pa_control_arxiv}
\end{document}